\documentclass[journal]{IEEEtran}
\ifCLASSINFOpdf
\else
\usepackage[dvips]{graphicx}
\fi

\usepackage{graphicx}
\usepackage{cite}
\usepackage{amsmath,amssymb,amsfonts}
\usepackage{algorithmic}
\usepackage{graphicx}
\usepackage{subfigure}
\usepackage{textcomp}
\usepackage{color,xcolor}
\usepackage{colortbl}
\usepackage{threeparttable}
\usepackage{url}
\usepackage{hyperref}
\usepackage{bm}
\usepackage{multirow}
\usepackage{balance}
\usepackage{bbm}
\usepackage{enumerate}

\begin{document}
\newcommand{\RNum}[1]{\uppercase\expandafter{\romannumeral #1\relax}}

\title{A Hybrid SFANC-FxNLMS Algorithm for Active Noise Control based on Deep Learning}
\author{Zhengding Luo, \IEEEmembership{Student Member, IEEE,} Dongyuan Shi, \IEEEmembership{Member, IEEE,} \\ and Woon-Seng Gan, \IEEEmembership{Senior Member, IEEE}
\thanks{\textit{(Corresponding author: Dongyuan Shi.)}}
\thanks{The authors are with the School of Electrical and Electronic Engineering, Nanyang Technological University, Singapore 639798, (e-mail: LUOZ0021@
e.ntu.edu.sg; dongyuan.shi@ntu.edu.sg; ewsgan@ntu.edu.sg). This paper has supplementary downloadable material available at \url{https://github.com/Luo-Zhengding/SFANC-FxNLMS-ANC-Algorithm-based-on-Deep-Learning}}
}
\maketitle

\begin{abstract}
The selective fixed-filter active noise control (SFANC) method selecting the best pre-trained control filters for various types of noise can achieve a fast response time. However, it may lead to large steady-state errors due to inaccurate filter selection and the lack of adaptability. In comparison, the filtered-X normalized least-mean-square (FxNLMS) algorithm can obtain lower steady-state errors through adaptive optimization. Nonetheless, its slow convergence has a detrimental effect on dynamic noise attenuation. Therefore, this paper proposes a hybrid SFANC-FxNLMS approach to overcome the adaptive algorithm's slow convergence and provide a better noise reduction level than the SFANC method. A lightweight one-dimensional convolutional neural network (1D CNN) is designed to automatically select the most suitable pre-trained control filter for each frame of the primary noise. Meanwhile, the FxNLMS algorithm continues to update the coefficients of the chosen pre-trained control filter at the sampling rate. Owing to the effective combination of the two algorithms, experimental results show that the hybrid SFANC-FxNLMS algorithm can achieve a rapid response time, a low noise reduction error, and a high degree of robustness.
\end{abstract}

\IEEEpeerreviewmaketitle

\section{Introduction}
\IEEEPARstart{A}{ctive} noise control (ANC) involves the generation of an anti-noise of equal amplitude and opposite phase to suppress the unwanted noise \cite{1,elliot1994active,kuo1999active,kajikawa2012recent,qiu2019introduction,zhang2019active,chen2021secondary}. Owing to its compact size and high efficiency in attenuating low frequency noises, ANC has been widely used in industrial operations and consumer products such as headphones \cite{liebich2018signal,rivera2017evaluation,2,chang2016listening,liebich2016active,shen2021wireless,coker2019survey}. Traditional ANC systems typically employ adaptive algorithms to attenuate noises \cite{3,guo2020convergence,yang2018frequency}. Filtered-X normalized least-mean-square (FxNLMS), being an adaptive ANC algorithm, not only can achieve low noise reduction errors through adaptive optimization but also does not require prior knowledge of the input data \cite{4,shi2020activeNormalized,lu2021survey}. However, due to the least mean square (LMS) based algorithms' inherent slow convergence and poor tracking ability \cite{26}, FxNLMS is less capable of dealing with rapidly varying or non-stationary noises. Its slow response to these noises may impact customers' perceptions of noise reduction effect \cite{22}.

In industrial applications and mature ANC products, fixed-filter ANC methods \cite{5,6} are adopted to tackle the problems of adaptive algorithms, where the control filter coefficients are fixed rather than adaptively updated to avoid slow convergence and instability. However, the coefficients of the fixed-filter algorithm are pre-trained for a specific noise type, resulting in the degradation of noise reduction performance for other types of noises \cite{11}. Therefore, for flexible selection of different pre-trained control filters in response to different noise types, Shi et al. \cite{7} proposed a selective fixed-filter active noise control (SFANC) method based on the frequency band matching. Nevertheless, several critical parameters of the method can only be determined through trials and errors. Also, its filter-selection module consumes extra computations in the real-time processor \cite{12}.

Thereby, automatic learning of the SFANC algorithm's critical parameters is a significant issue that would promote its uses in real-world scenarios and industrial products. Under this consideration, deep learning techniques, particularly convolutional neural networks (CNNs), achieving significant success in classification \cite{8,21,sikora2021influence,zhang2021deep} appear to be an excellent approach for improving SFANC. By replacing the filter-selection module with a CNN model, the SFANC algorithm can automatically learn its parameters from noise datasets and select the best control filter given different noise types without resorting to extra-human efforts \cite{9}. Additionally, a CNN model implemented on a co-processor can decouple the computational load from the real-time noise controller \cite{9}. Though its superiority on response speed has been demonstrated, the CNN-based SFANC algorithm is limited by some wrong noise classifications and the lack of adaptive optimization, which may result in large steady-state errors.

As discussed above, using either SFANC or FxNLMS is difficult to constantly obtain satisfactory noise reduction performance for various noises. The SFANC algorithm and FxNLMS algorithm can be effectively combined to overcome the limitations of the individual ANC algorithms. Therefore, a hybrid SFANC-FxNLMS algorithm is proposed in this paper. In the proposed algorithm, a lightweight one-dimensional CNN (1D CNN) implemented on a co-processor can automatically select the most suitable control filter for different noise types. For each frame of the noise waveform, the control filter coefficients are decided by the 1D CNN and continue to be updated adaptively via the FxNLMS algorithm at the sampling rate. Simulation results show that the proposed algorithm not only achieves faster noise reduction responses and lower steady-state errors but also exhibits good tracking and robustness. Thus, it can be used for attenuating dynamic noises such as traffic noises and urban noises etc.

The remainder of this paper is organized as follows: Section \RNum{2} introduces the hybrid SFANC-FxNLMS algorithm in detail. Section \RNum{3} presents a series of numerical simulations to evaluate the effectiveness of the proposed algorithm. Finally, the conclusion is given in Section \RNum{4}.

\begin{figure}[tp]
\setlength{\abovecaptionskip}{0.cm}
\setlength{\belowcaptionskip}{-0.cm}
\centering
\centerline{\includegraphics[height=5cm,width=\linewidth]{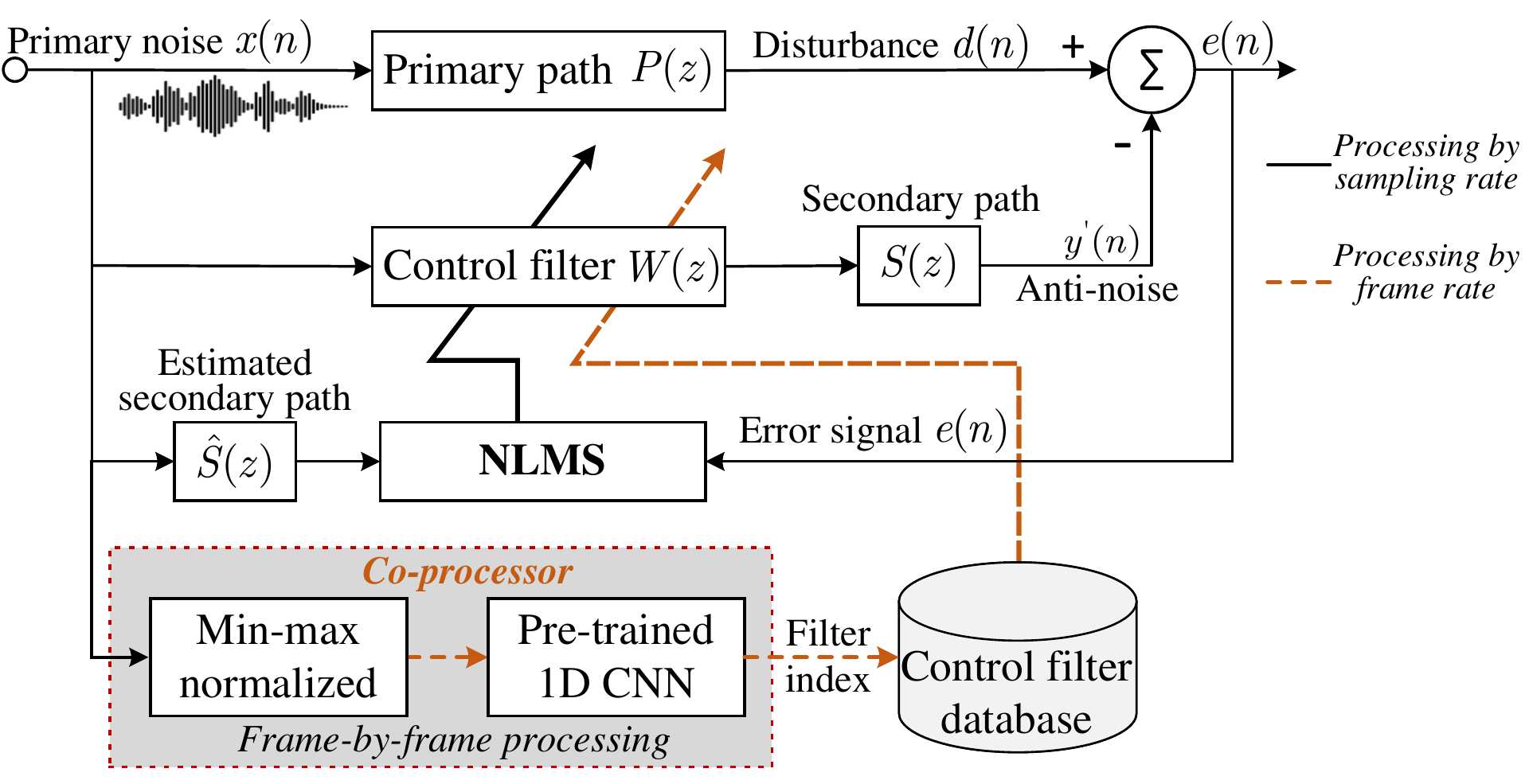}}
\caption{Block diagram of the hybrid SFANC-FxNLMS algorithm. For each frame, the most suitable pre-trained control filter is selected by the 1D CNN operated in a co-processor. Subsequently, FxNLMS continues to update the coefficients of the chosen pre-trained control filter at the sampling rate.}
\label{Fig 1}\vspace*{-0.3cm}
\end{figure}

\section{Hybrid SFANC-FxNLMS Algorithm}
The overall framework of the hybrid SFANC-FxNLMS algorithm is presented in Fig.~\ref{Fig 1}. Although learning acoustic models directly from the raw waveform data is challenging \cite{10}, the input of the proposed ANC system is raw waveform instead of frequency spectrogram.\vspace*{-0.3cm}

\subsection{A concise explanation of SFANC}
\begin{figure}[tp]
\setlength{\abovecaptionskip}{0.cm}
\setlength{\belowcaptionskip}{-0.cm}
    \centering
    \includegraphics[width=0.75\linewidth]{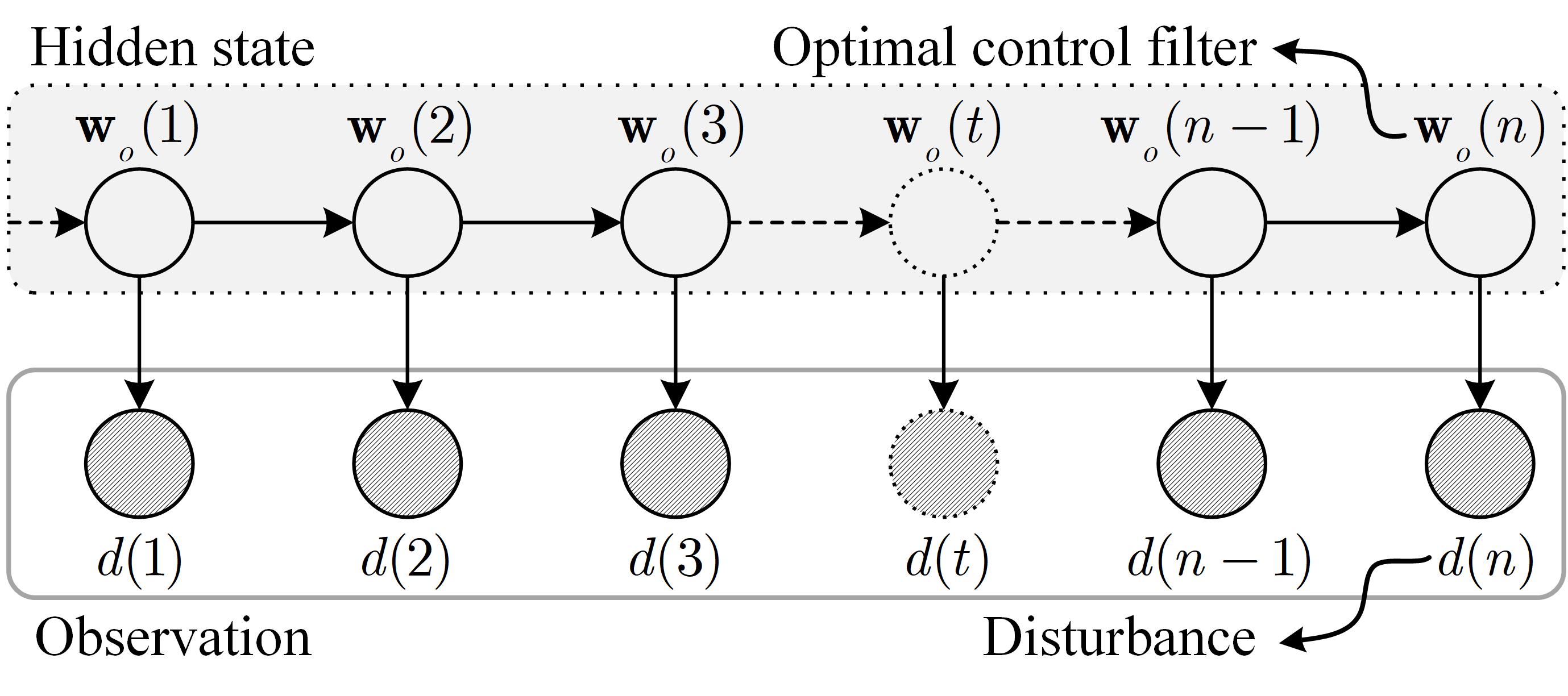}
    \caption{Markov model of the active noise control progress}
    \label{Fig 2}
\end{figure}

An ANC progress can be abstracted as a first-order Markov chain \cite{23,24}, as shown in Fig.~\ref{Fig 2}, where $\mathbf{w}_o(n)$ represents the optimal control filter to deal with the disturbance $d(n)$. If we assumed that the optimal control filter has $C$ discrete states as $\{\mathbf{w}_i\}^C_{i=1}$, the control filter selection process can be considered as determining the best control filter from a pre-trained filter set $\{\mathbf{w}_i\}^C_{i=1}$.
Hence, the SFANC strategy can be summarized as follows:
\begin{equation}\label{eq_s1}
    \mathbf{w}_o =  \underset{\mathbf{w}\in\{\mathbf{w}_i\}^C_{i=1}}{\mathrm{argmin}}~\mathbb{E}\bigg\{\left[d(n)-\mathbf{x}^\mathrm{T}(n)\mathbf{w}(n)\ast s(n)\right]^2\bigg\},
\end{equation}
where $\text{argmin}(\cdot)$ operator returns the input value for minimum output; $\ast$ , $\mathbf{x}(n)$, and $\mathbf{s}(n)$ represent the linear convolution, the reference signal, and the impulse response of the secondary path, respectively. The reference signal is assumed to be the same as the primary noise.

In practice, $d(n)$ is typically regarded as the linear combination of $\mathbf{x}(n)$. Thus, \eqref{eq_s1} is equivalent to 
\begin{equation}\label{eq_s2}
    \mathbf{w}_{o} = \underset{\mathbf{w}\in\{\mathbf{w}_i\}^C_{i=1}}{\mathrm{argmax}} P\left[\mathbf{w}|d(n)\right] =  \underset{\mathbf{w}\in\{\mathbf{w}_i\}^C_{i=1}}{\mathrm{argmax}} P\left[\mathbf{w}|\mathbf{x}(n)\right],
\end{equation}
which refers to the process of selecting a control filter from the pre-trained filter set that maximizes its posterior probability in the presence of reference signal $\mathbf{x}(n)$.
Moreover, according to Bayes' theorem~\cite{kay1993fundamentals}, the posterior probability in \eqref{eq_s2} can be replaced with a conditional probability as 
\begin{equation}
    \mathbf{w}_{o} =  \underset{\mathbf{w}\in\{\mathbf{w}_i\}^C_{i=1}}{\mathrm{argmax}} P\left[\mathbf{x}(n)|\mathbf{w}\right],
\end{equation}
which predicts the most suitable control filter straight from the reference signal $\mathbf{x}(n)$.

To achieve a satisfactory noise reduction for the SFANC algorithm, we should find a classifier model $\hat{P}\left[\mathbf{x}(n)|\mathbf{w},\Theta\right]$ to approximate ${P}\left[\mathbf{x}(n)|\mathbf{w}\right]$ from the pre-recorded sampling set $\{\mathbf{x}^{j}(n), \mathbf{w}^{j}\}^{N}_j$. The $\Theta$ denotes the parameters of the classifier and can be obtained through maximum likelihood estimation (MLE) \cite{25} as
\begin{equation}
    \Theta = \underset{\Theta}{\mathrm{argmax}}\frac{1}{N}\sum^{N}_{j=1}\log{\hat{P}\left[\mathbf{x}^{j}(n)|\mathbf{w}^{j},\Theta\right]}.
\end{equation}
Hence, we can utilize the deep learning approach to extract this classifier model from the training set $\{\mathbf{x}^{j}(n), \mathbf{w}^{j}\}^{N}_j$.\vspace*{-0.2cm}

\begin{figure}[tp]
\setlength{\abovecaptionskip}{0.cm}
\setlength{\belowcaptionskip}{-0.cm}
\centering
\centerline{\includegraphics[height=2.5cm,width=0.75\linewidth]{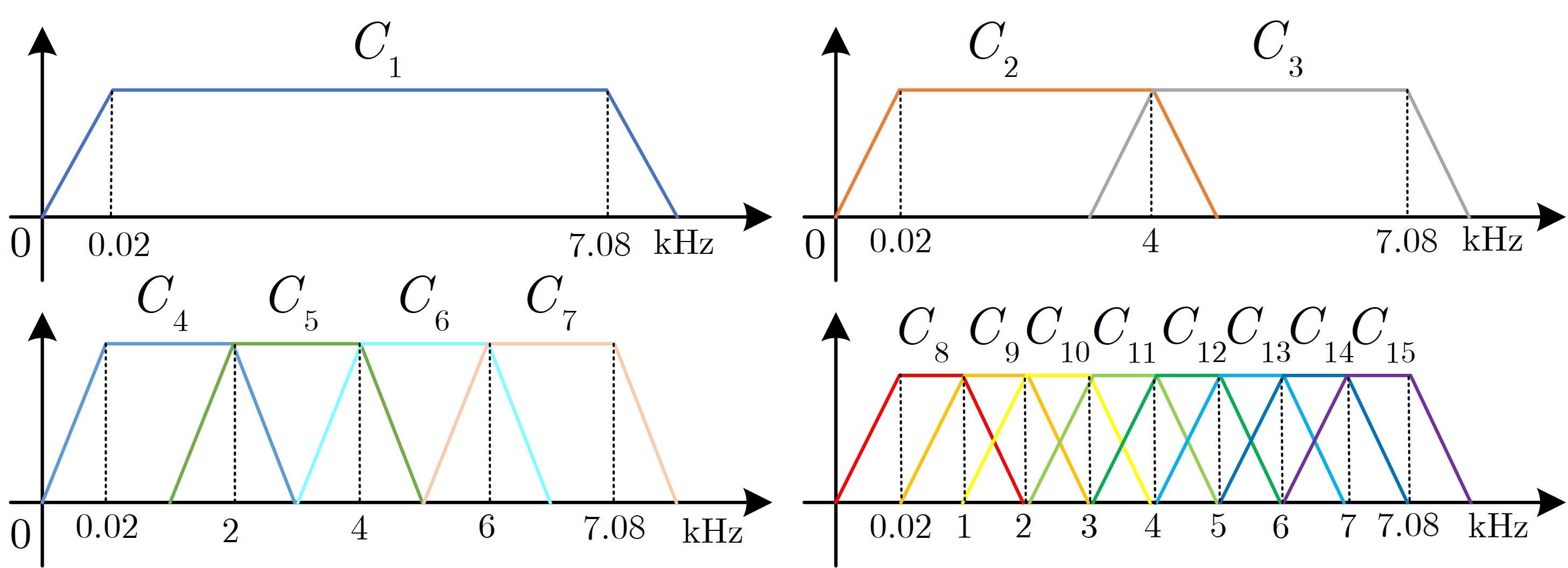}}
\caption{The frequency bands of the broadband noises in pre-training stage.}
\label{Fig 3}\vspace*{-0.3cm}
\end{figure}

\subsection{CNN-based SFANC algorithm}
Motivated by the work \cite{9}, this paper develops a 1D CNN capable of dealing with noise waveforms end-to-end. The input of the network is the normalized noise waveform with a 1-second duration. The min-max operation is defined as:
\begin{equation}
\hat{x}(n)=\frac{x(n)}{\max [\mathbf{x}(n)]-\min [\mathbf{x}(n)]},
\end{equation}
where $\max [\cdot]$ and $\min [\cdot]$ obtain the maximum and minimum value of the input vector $\mathbf{x}(n)$. It aims to rescale the input range into $(-1,1)$ and retain the signal’s negative part that contains phase information.

\subsubsection{Pre-training control filters}
The primary and secondary paths used in the training stage of the control filters are band-pass filters with a frequency range of 20Hz-7,980Hz. Broadband noises with $15$ frequency ranges shown in Fig.~\ref{Fig 3} are used to pre-train $15$ control filters. The filtered reference least mean square (FxLMS) algorithm is adopted to obtain the optimal control filters for these broadband noises due to its low computational complexity~\cite{yang2020stochastic}. Subsequently, these $15$ pre-trained control filters are saved in the control filter database.

\subsubsection{The Proposed 1D CNN}
\begin{figure}[tp]
\setlength{\abovecaptionskip}{0.cm}
\setlength{\belowcaptionskip}{-0.cm}
\centering
\centerline{\includegraphics[height=5cm,width=0.5\linewidth]{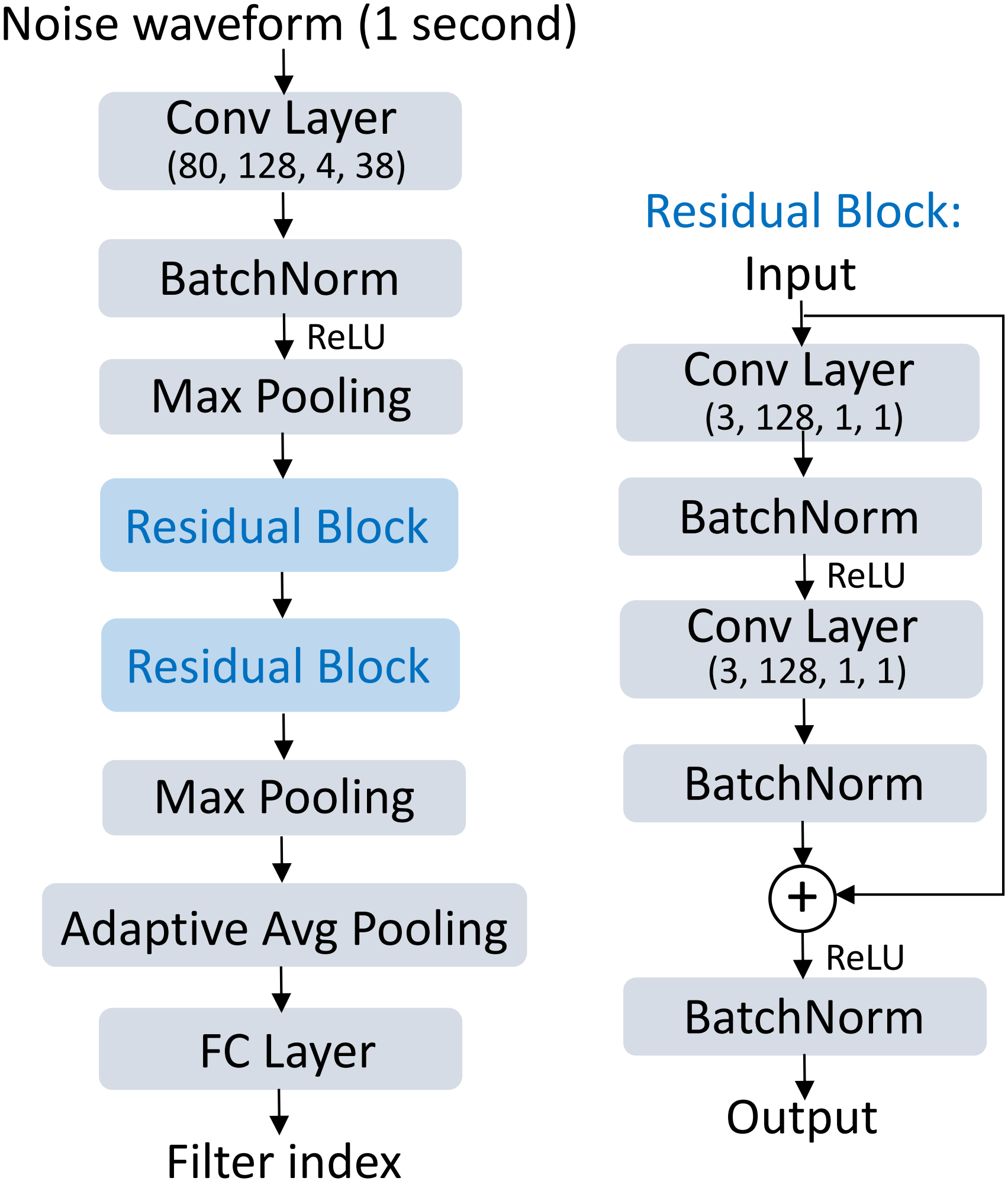}}
\caption{Architecture of the 1D CNN used to select the most suitable pre-trained control filter for each frame of the noise waveform. The configuration of convolution layer is denoted as: (kernel size, channels, stride, padding).}
\label{Fig 4}\vspace*{-0.3cm}
\end{figure}

The detailed architecture of the 1D CNN is illustrated in Fig.~\ref{Fig 4}. Every residual block in the network is composed of two convolutional layers, subsequent batch normalization \cite{19}, and ReLU non-linearity \cite{20}. Note that a shortcut connection is adopted to add the input with the output in each residual block since residual architecture is demonstrated easy to be optimized \cite{14}. Additionally, the network uses a broad receptive field (RF) in the first convolutional layer and narrow RFs in the rest convolutional layers to fully exploit both global and local information.

In order to train the 1D CNN model, we generated $80,000$ broadband noise tracks with various frequency bands, amplitudes, and background noise levels at random. Each track has a duration of 1 second. These noise tracks are taken as primary noises to generate disturbances. The class label of a noise track corresponds to the index of the control filter that achieves the best noise reduction performance on the disturbance.

\subsection{The FxNLMS algorithm}
The FxNLMS algorithm is a conventional adaptive ANC algorithm \cite{15}, where the error signal can be obtained from:
\begin{equation}
e(n) = d(n)-\mathbf{x}^{\mathrm{T}}(n) \mathbf{w}(n) \ast s(n),
\end{equation}
where $\mathbf{w}(n)$, $(\cdot)^{\mathrm{T}}$ and $\ast$ represent the control filter, the transposed operation and linear convolution, respectively. The weight update equation of control filter is given by:
\begin{equation}\label{eq_lms}
\begin{aligned}
\mathbf{w}(n+1) &=\mathbf{w}(n)+\mu \mathbf{r}(n) e(n) \Gamma_{k}^{-1}, \\
\mathbf{r}(n) &=\hat{\mathbf{s}}(n) * \mathbf{x}(n), \\
\Gamma_{k} &=\beta+\mathbf{r}^{T}(n) \mathbf{r}(n),
\end{aligned}
\end{equation}
in which, $\mu$ denotes the step size. $\mathbf{r}(n)$ is the filtered reference signal generated by passing the reference signal $\mathbf{x}(n)$ through the estimated secondary path $\hat{\mathbf{s}}(n)$. Additionally, $\beta$ is a small constant that avoids dividing by zero in the denominator.

\subsection{Hybrid SFANC-FxNLMS Algorithm}
As illustrated in Fig.~\ref{Fig 1}, the primary noise is sampled at a rate of $16,000$Hz, while each data frame is made up of $16,000$ samples ($1$ second duration data). Throughout the control process, the co-processor (e.g., a mobile phone) utilizes the pre-trained 1D CNN to output the index of the most appropriate control filter for each frame and sends it to the real-time controller. When the received filter index differs from the current index, the real-time controller will swap the control filter. Simultaneously, the real-time controller continues to update the control filter's coefficients via \eqref{eq_lms} at each sample. Notably, the co-processor operates at the frame rate, whereas the real-time controller performs at the sampling rate.

The hybrid SFANC-FxNLMS algorithm can achieve faster responses to different noises with the assistance of the SFANC. Meanwhile, the FxNLMS algorithm continues to update the control filter's coefficients in the SFANC-FxNLMS algorithm, which contributes to obtain lower noise reduction errors.\vspace*{-0.3cm}

\begin{table}[tp]
\setlength{\abovecaptionskip}{0.cm}
\setlength{\belowcaptionskip}{-0.cm}
\caption{Comparisons of different 1D networks}
\begin{center}
\begin{tabular}{l|c|c|c}
\hline
\textbf{Network} & \textbf{Test Accuracy} & \textbf{\#Layers} & \textbf{\#Parameters} \\
\hline
Proposed Network & \textbf{99.35$\%$} & \textbf{6} & \textbf{0.21M} \\

M3 Network & 99.15$\%$ & 3 & 0.22M \\

M5 Network & 99.10$\%$ & 5 & 0.56M \\

M11 Network & 99.15$\%$ & 11 & 1.79M \\

M18 Network & 98.50$\%$ & 18 & 3.69M \\
M34-res Network & 98.35$\%$ & 34 & 3.99M \\
\hline
\end{tabular}\vspace*{-0.6cm}
\label{Table 1}
\end{center}
\end{table}

\section{Experiments}
In the experiments, we trained the 1D CNN model on a synthesized noise dataset and then evaluated the hybrid SFANC-FxNLMS algorithm on real-record noises. The primary path and secondary path are chosen as band-pass filters. Moreover, the step size of the FxNLMS algorithm is set to $0.002$, and the control filter length is $1,024$ taps.\vspace*{-0.4cm}

\subsection{Training and testing of 1D CNN}
The synthetic noise dataset was subdivided into three subsets: $80,000$ noise tracks for training, $2,000$ noise tracks for validation, and $2,000$ noise tracks for testing. During training, the Adam algorithm~\cite{17} was used for optimization. The training epoch was set at $30$. To avoid exploding or vanishing gradients, the glorot initialization~\cite{16} was used. To avoid overfitting due to the scarcity of training data, the 1D network was built as a light-weighted network with weights all subjected to $\ell_{2}$ regularization with a coefficient of $0.0001$.

The proposed 1D CNN is compared to several different 1D networks proposed in~\cite{10} utilizing raw acoustic waveforms: M3, M5, M11, M18, and M34-res. Table~\ref{Table 1} summarizes the comparison results. The number of network layers includes the number of convolutional layers and fully connected layers. Among these networks, the proposed 1D network obtains the highest classification accuracy of $99.35\%$ on the test dataset, which means it can provide the most suitable pre-trained control filters for different noises. Also, the proposed 1D network is lightweight, utilizing the fewest network parameters possible due to its efficient architecture.\vspace*{-0.2cm}\vspace*{-0.2cm}

\subsection{Real-world noise cancellation}
This section presents using the SFANC, FxNLMS, and hybrid SFANC-FxNLMS algorithm to attenuate real-record noises. The real-world noises are not included in the training datasets. We created two distinct noises by cascading and mixing an aircraft noise with a frequency range of 50Hz-14,000Hz and a traffic noise with a frequency range of 40Hz–1,400Hz.

\subsubsection{Cascaded noise attenuation}
\begin{figure}[tp]
\setlength{\abovecaptionskip}{0.cm}
\setlength{\belowcaptionskip}{-0.cm}
\centering
\subfigure{
\includegraphics[width=0.44\linewidth]{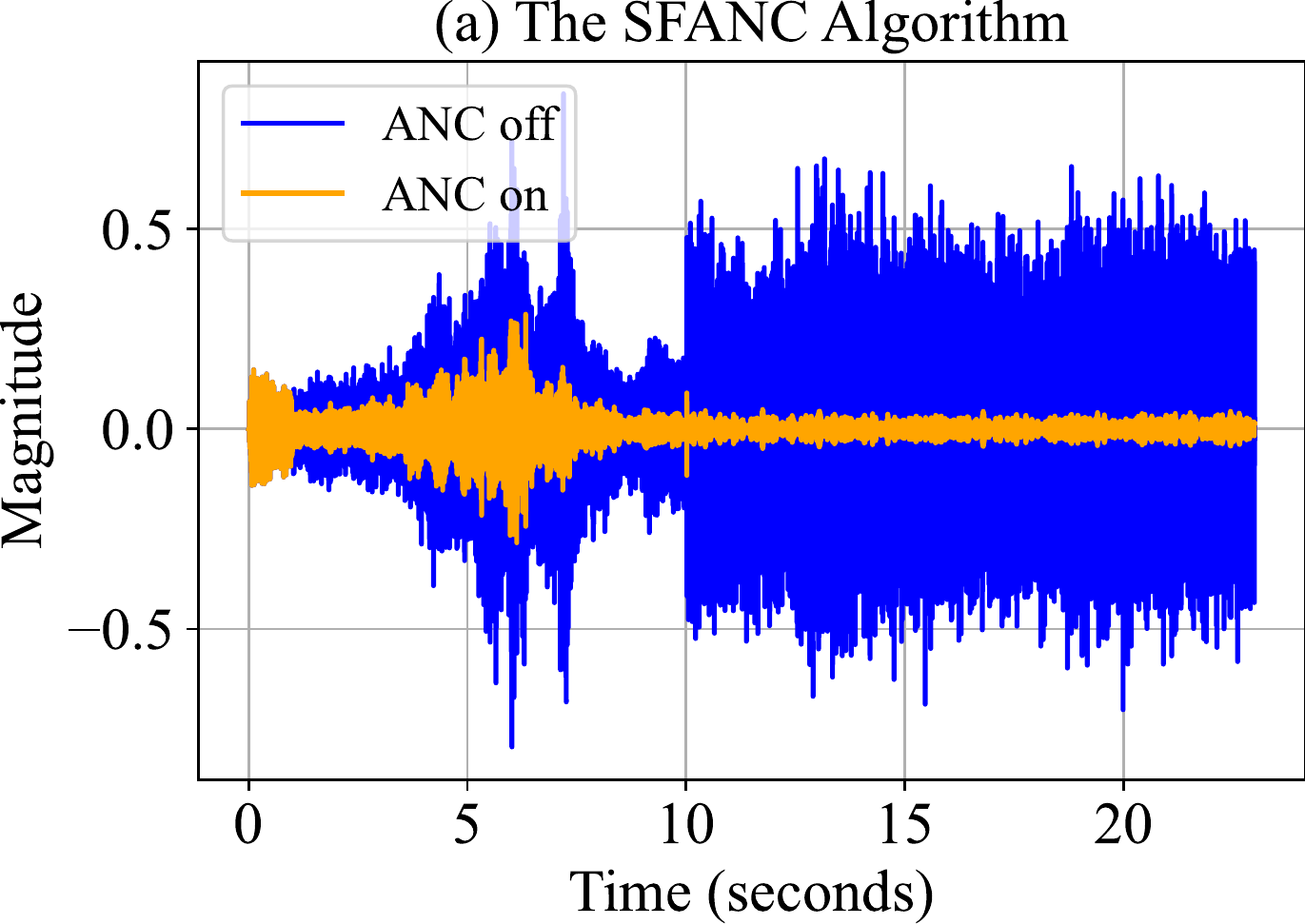}
}
\subfigure{
\includegraphics[width=0.44\linewidth]{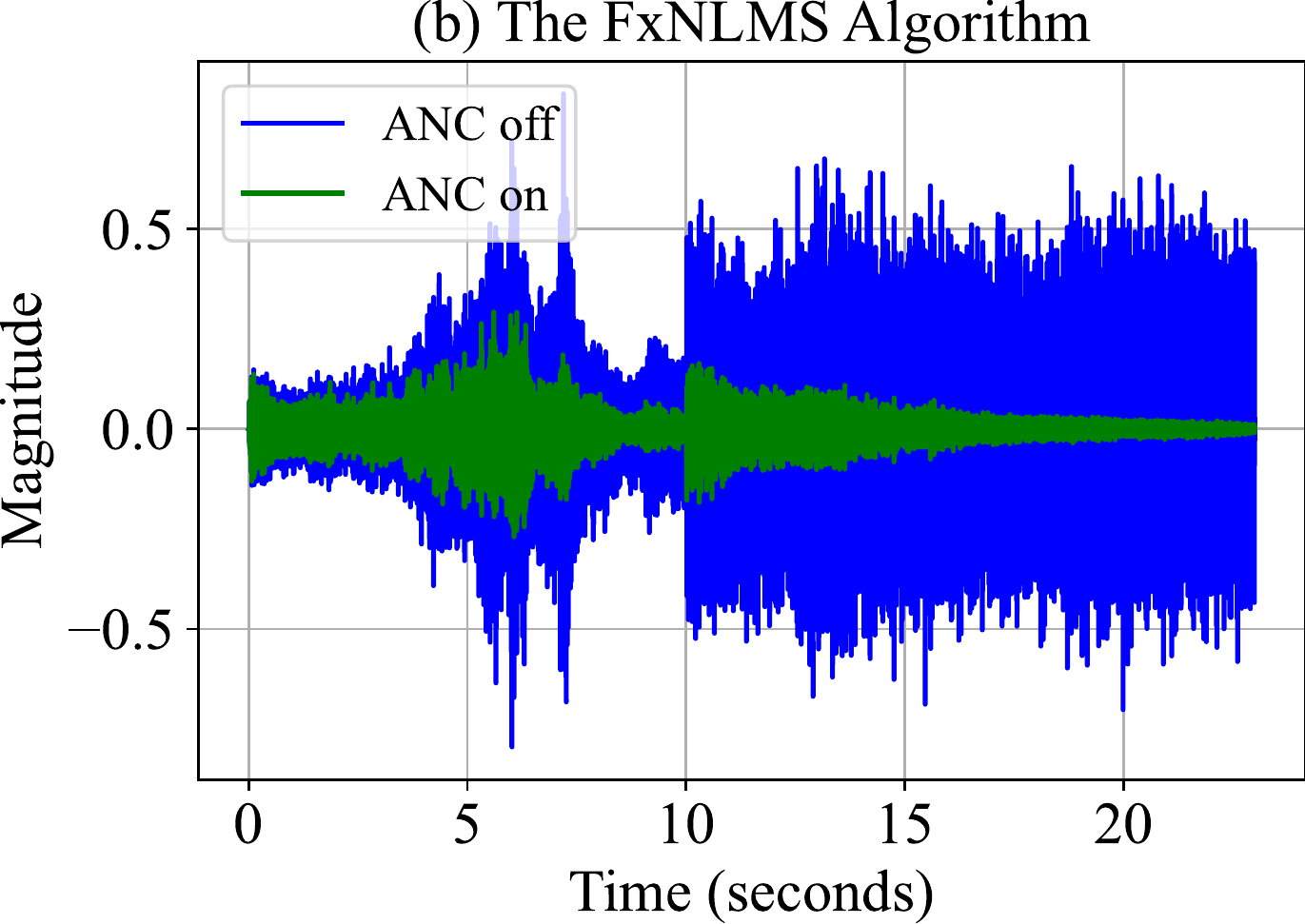}
}
\subfigure{
\includegraphics[width=0.44\linewidth]{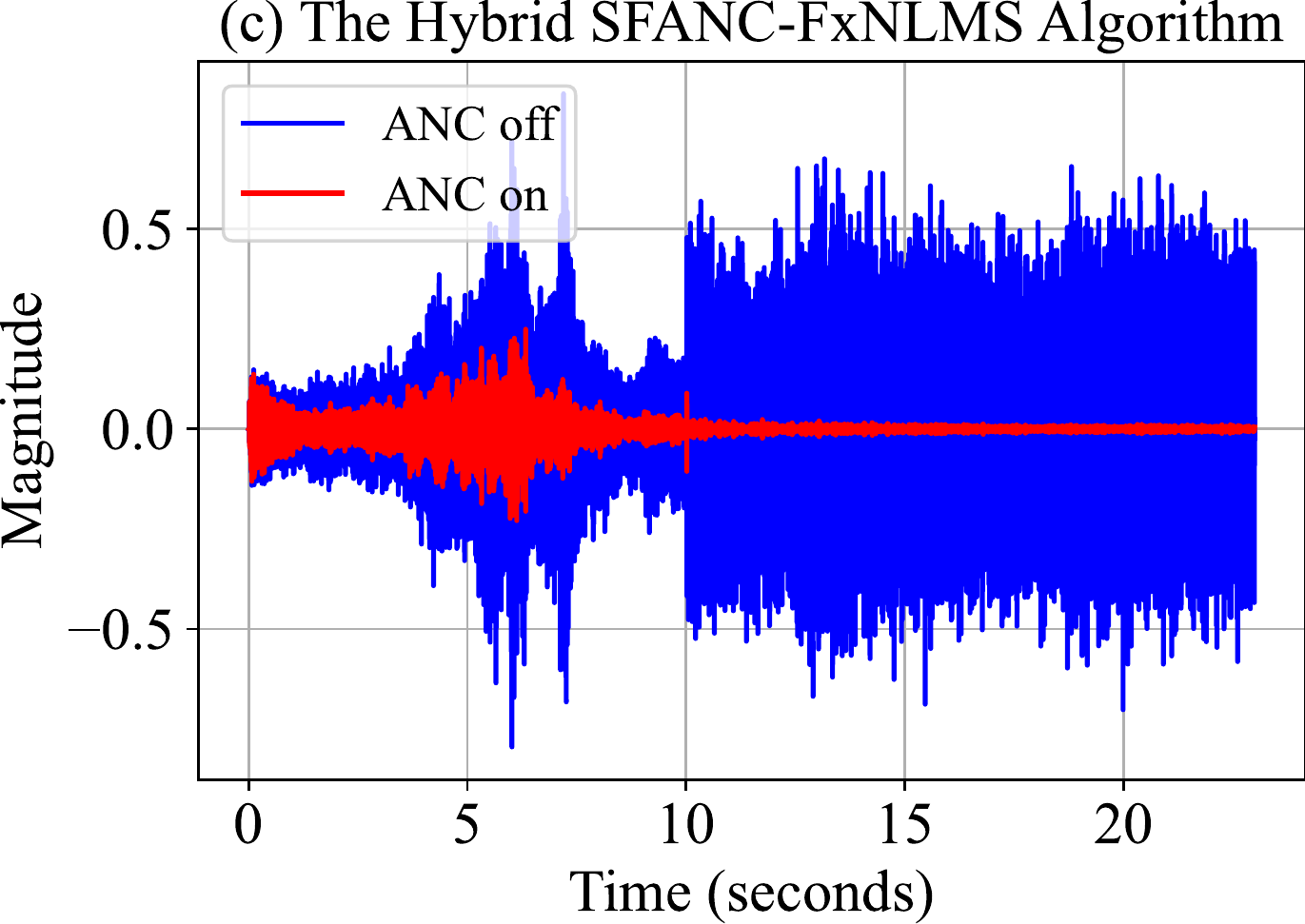}
}
\subfigure{
\includegraphics[width=0.44\linewidth]{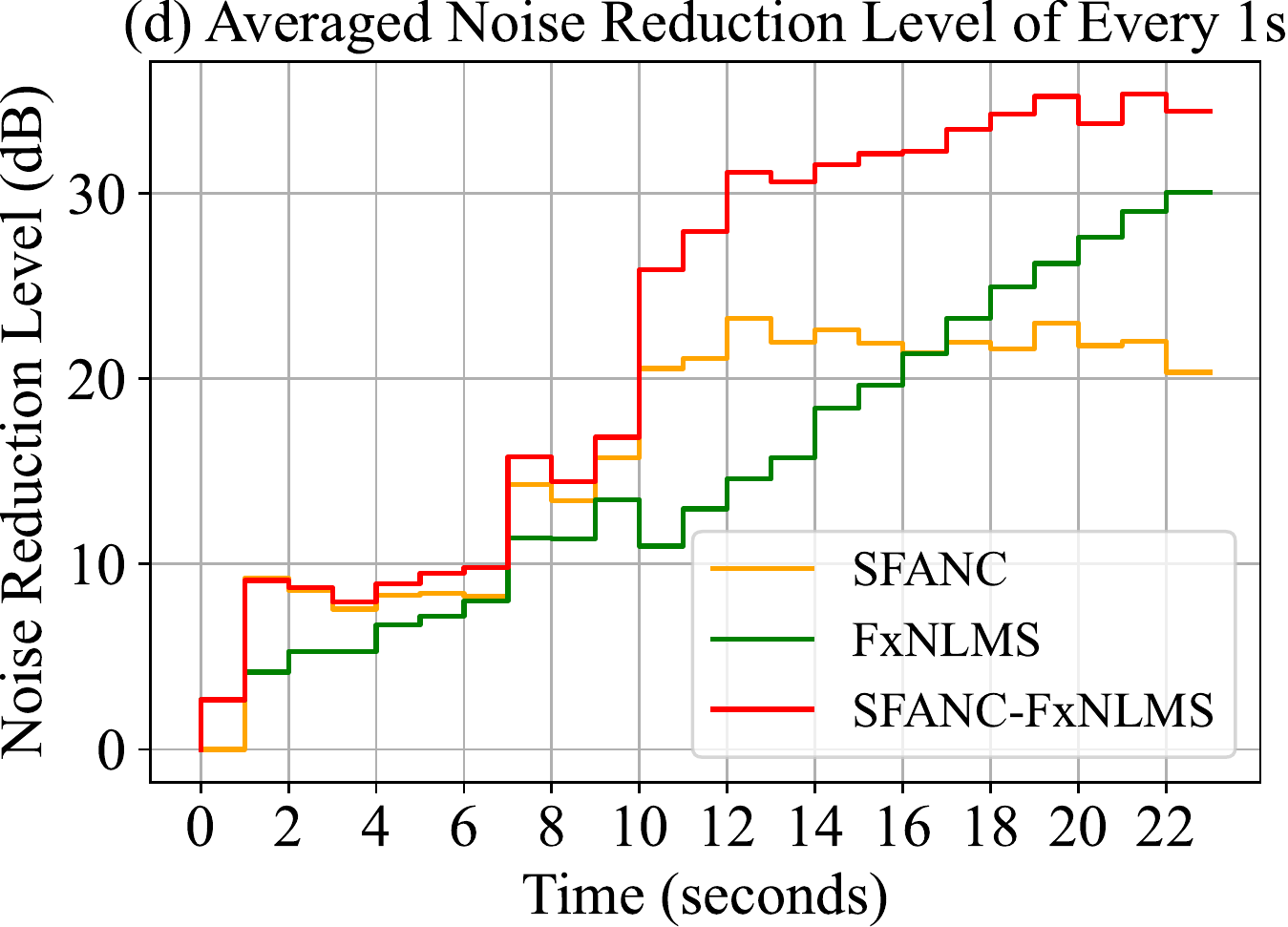}
}
\caption{(a)-(c): Error signals of different ANC algorithms, (d): Averaged noise reduction level of every 1 second, on the cascaded noise.}
\label{Fig 5}\vspace*{-0.3cm}
\end{figure}

The varying noise in this simulation is provided by cascading aircraft noise and traffic noise. The noise reduction results using different ANC methods on the cascaded noise are shown in Fig.~\ref{Fig 5}.

From the result, we can observe that the hybrid SFANC-FxNLMS algorithm can track and respond quickly to different parts of the varying noise. During the noise reduction process, the hybrid SFANC-FxNLMS algorithm consistently outperforms the SFANC and FxNLMS algorithms. In particular, during $10$s-$11$s, the averaged noise reduction level achieved by the hybrid SFANC-FxNLMS algorithm is 6dB and 15dB more than that of SFANC and FxNLMS, respectively. Noticeably, it is the transition period from one noise to another. Therefore, even when the frequency band and amplitude of the noise change, the hybrid SFANC-FxNLMS algorithm can still attenuate it.

\begin{figure}[tp]
\setlength{\abovecaptionskip}{0.cm}
\setlength{\belowcaptionskip}{-0.cm}
\centering
\subfigure{
\includegraphics[width=0.44\linewidth]{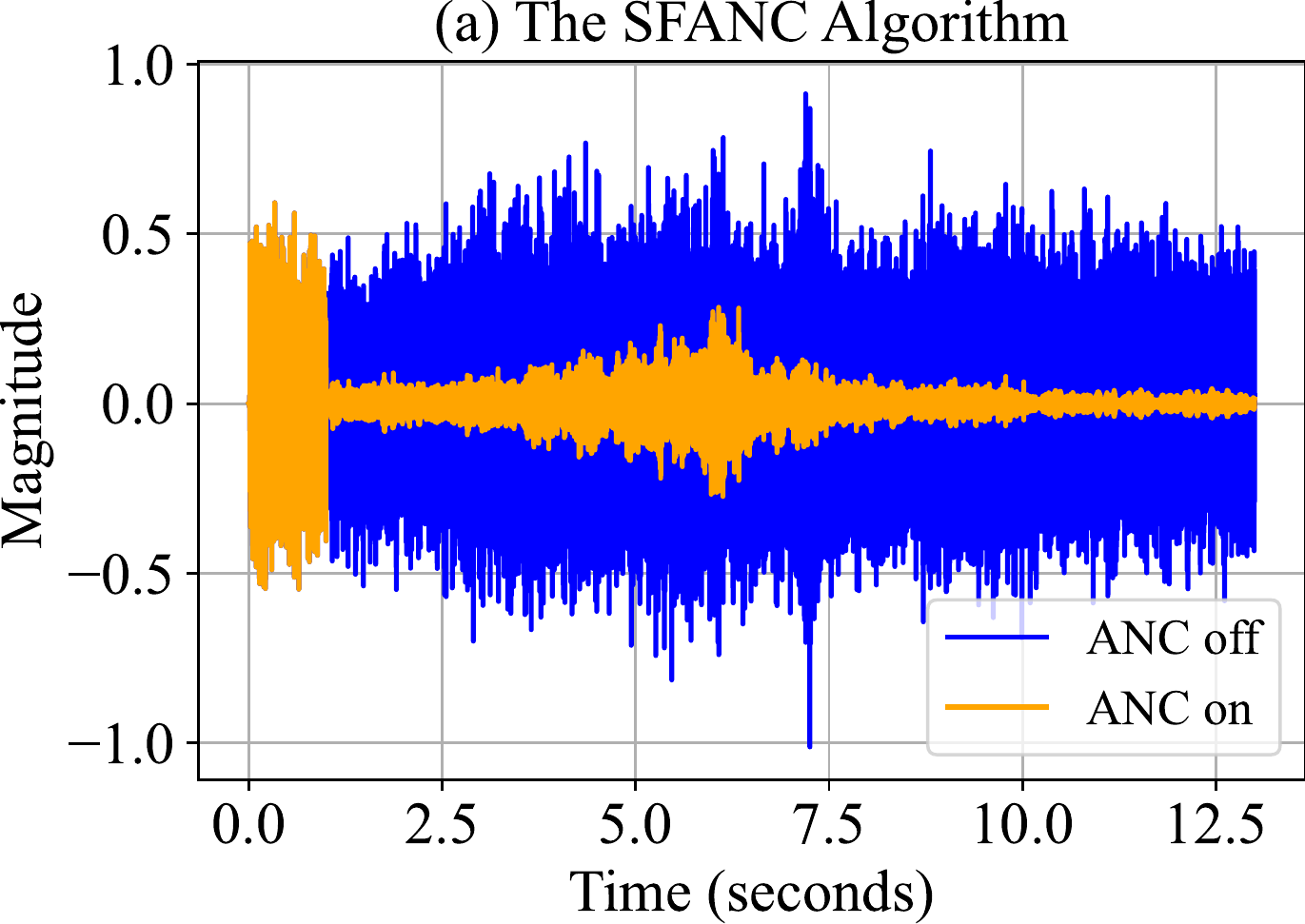}
}
\subfigure{
\includegraphics[width=0.44\linewidth]{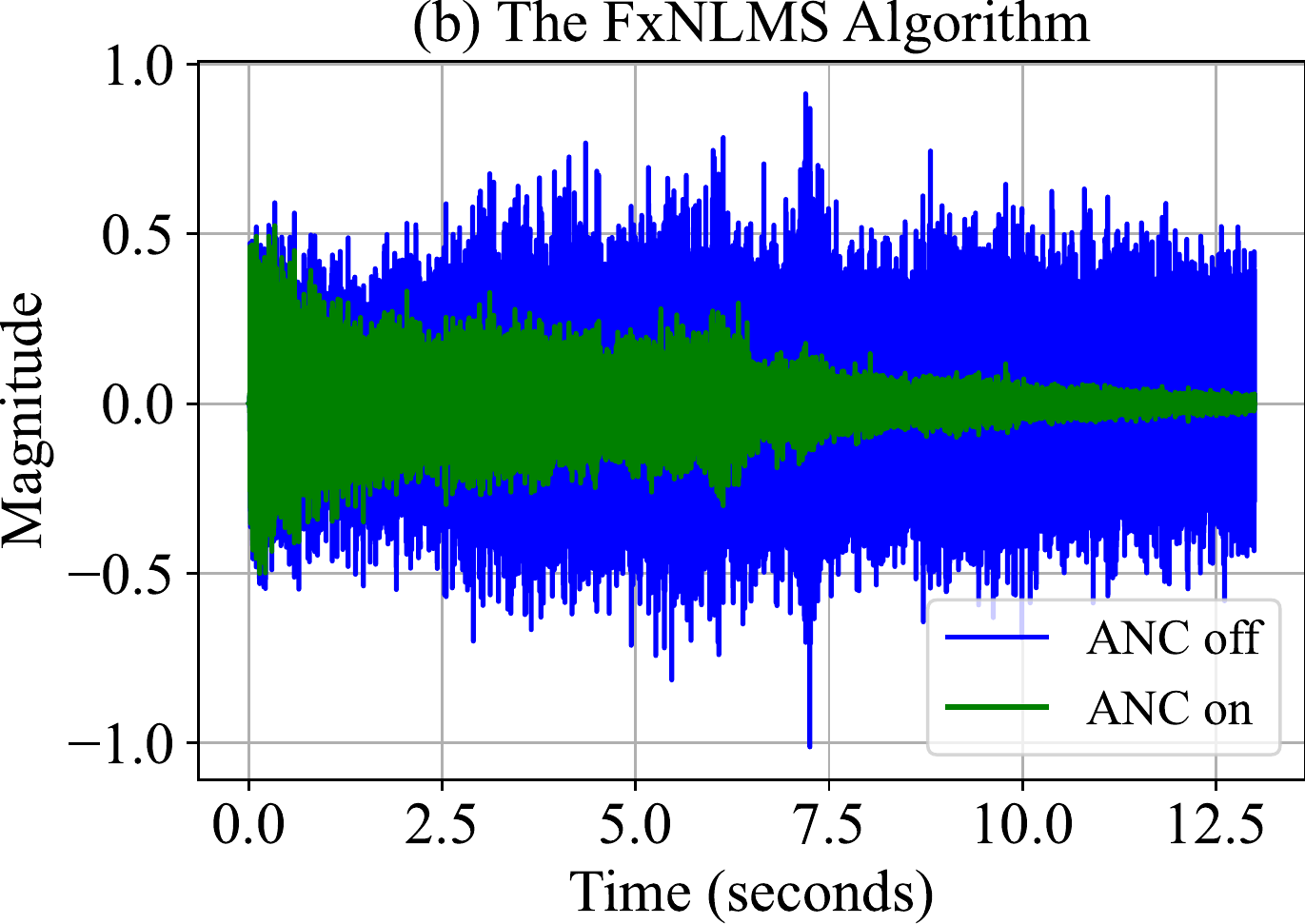}
}
\subfigure{
\includegraphics[width=0.44\linewidth]{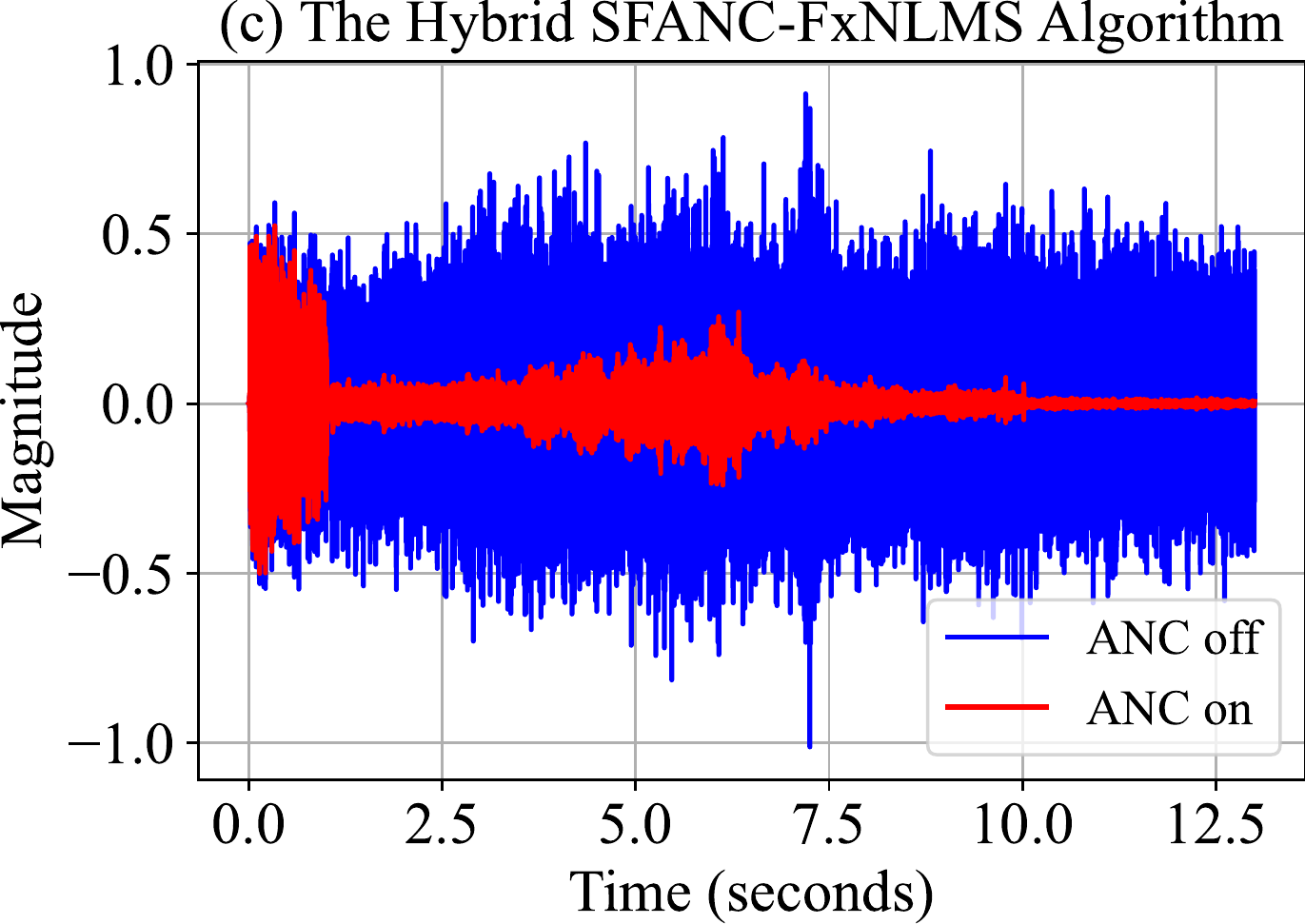}
}
\subfigure{
\includegraphics[width=0.44\linewidth]{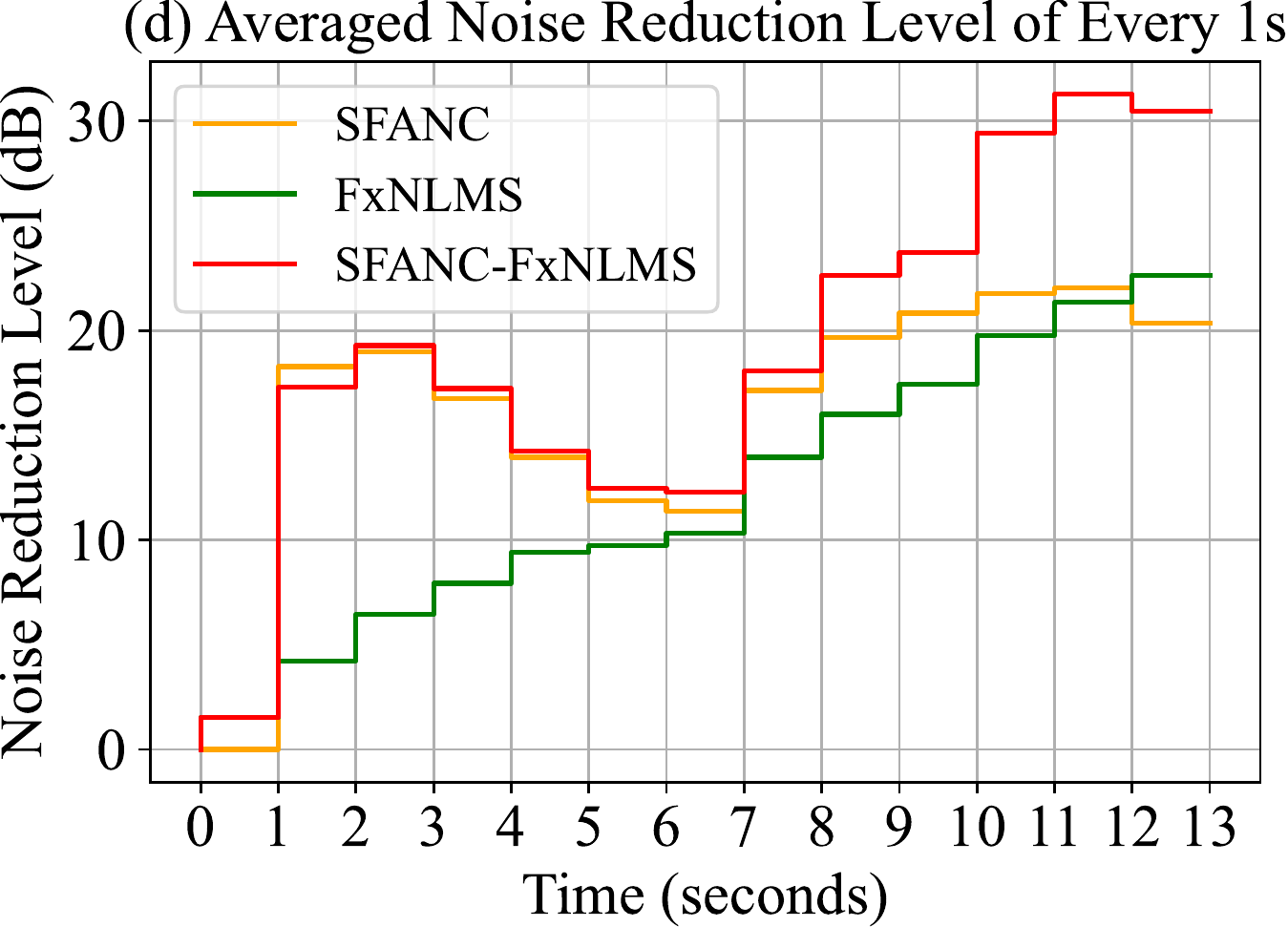}
}
\caption{(a)-(c): Error signals of different ANC algorithms, (d): Averaged noise reduction level of every 1 second, on the mixed noise.}
\label{Fig 6}\vspace*{-0.4cm}
\end{figure}

\subsubsection{Mixed noise attenuation}
In addition to cascading, we added the aircraft noise and traffic noise together to get a mixed noise. Fig.~\ref{Fig 6} depicts the performance of various ANC algorithms on the mixed noise. The SFANC algorithm responds to mixed noise much faster than the FxNLMS algorithm, as illustrated in the figure. For example, the SFANC method achieves a noise reduction level of about $18$dB after $1$s, whereas the FxNLMS algorithm takes more than $10$s to achieve the same level. However, after convergence during $12$s-$13$s, the FxNLMS algorithm achieves a higher noise reduction level than the SFANC method.

Also, it is noticeable that the SFANC algorithm is incapable of dealing with new noise during the first $1$s, since it only updates the control filter index after the first $1$s. However, owing to combining with FxNLMS, the hybrid SFANC-FxNLMS algorithm can slightly attenuate the noise during the first $1$s. Furthermore, we can see that the averaged noise reduction level of hybrid SFANC-FxNLMS algorithm is about $10$dB higher than that of SFANC and FxNLMS algorithms in $12$s-$13$s. Therefore, the results on the mixed noise also confirm the superiority of the hybrid SFANC-FxNLMS algorithm over the SFANC and FxNLMS algorithm.\vspace*{-0.2cm}

\begin{figure}[tp]
\setlength{\abovecaptionskip}{0.cm}
\setlength{\belowcaptionskip}{-0.cm}
\centering
\subfigure{
\includegraphics[width=0.44\linewidth]{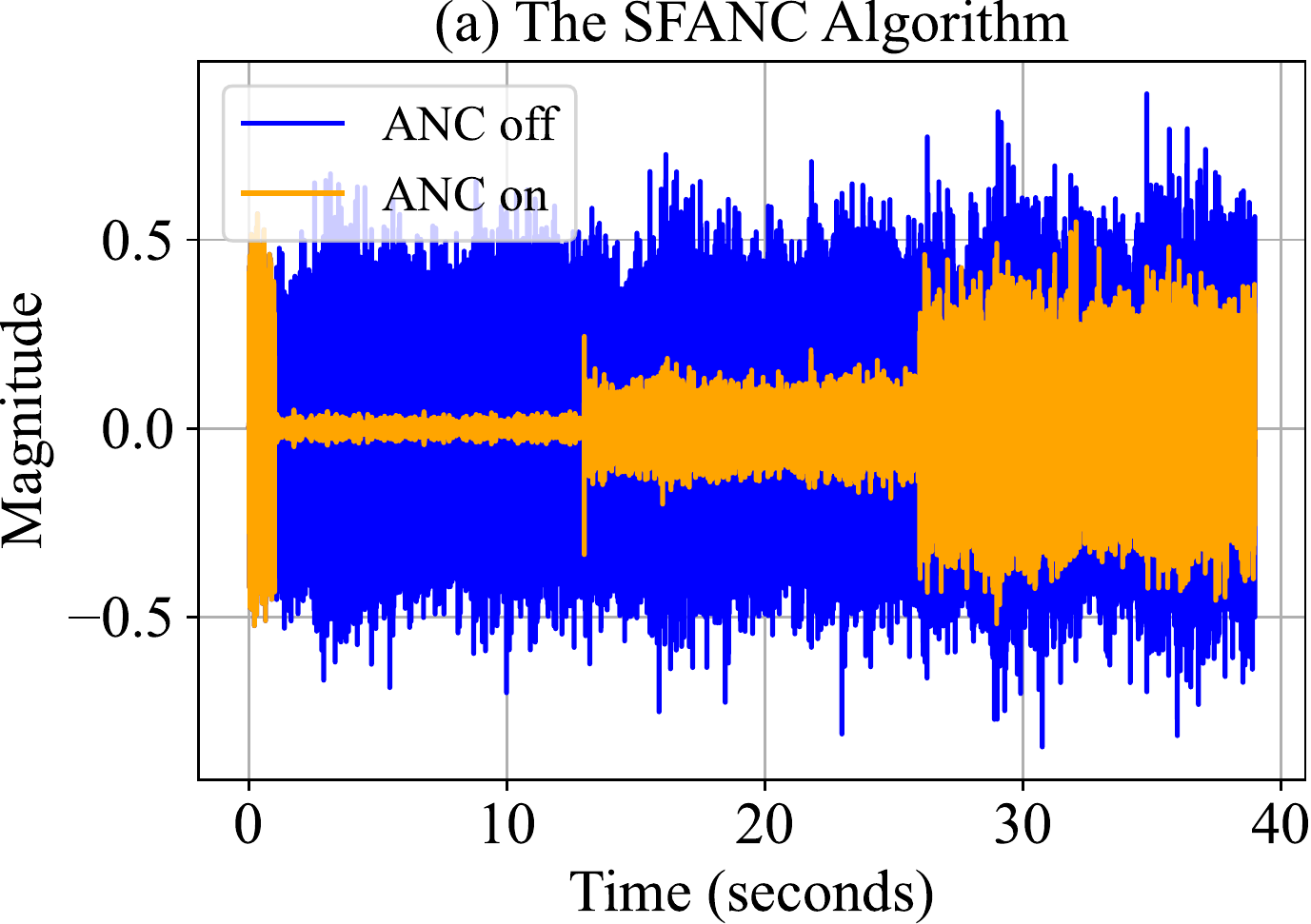}
}
\subfigure{
\includegraphics[width=0.44\linewidth]{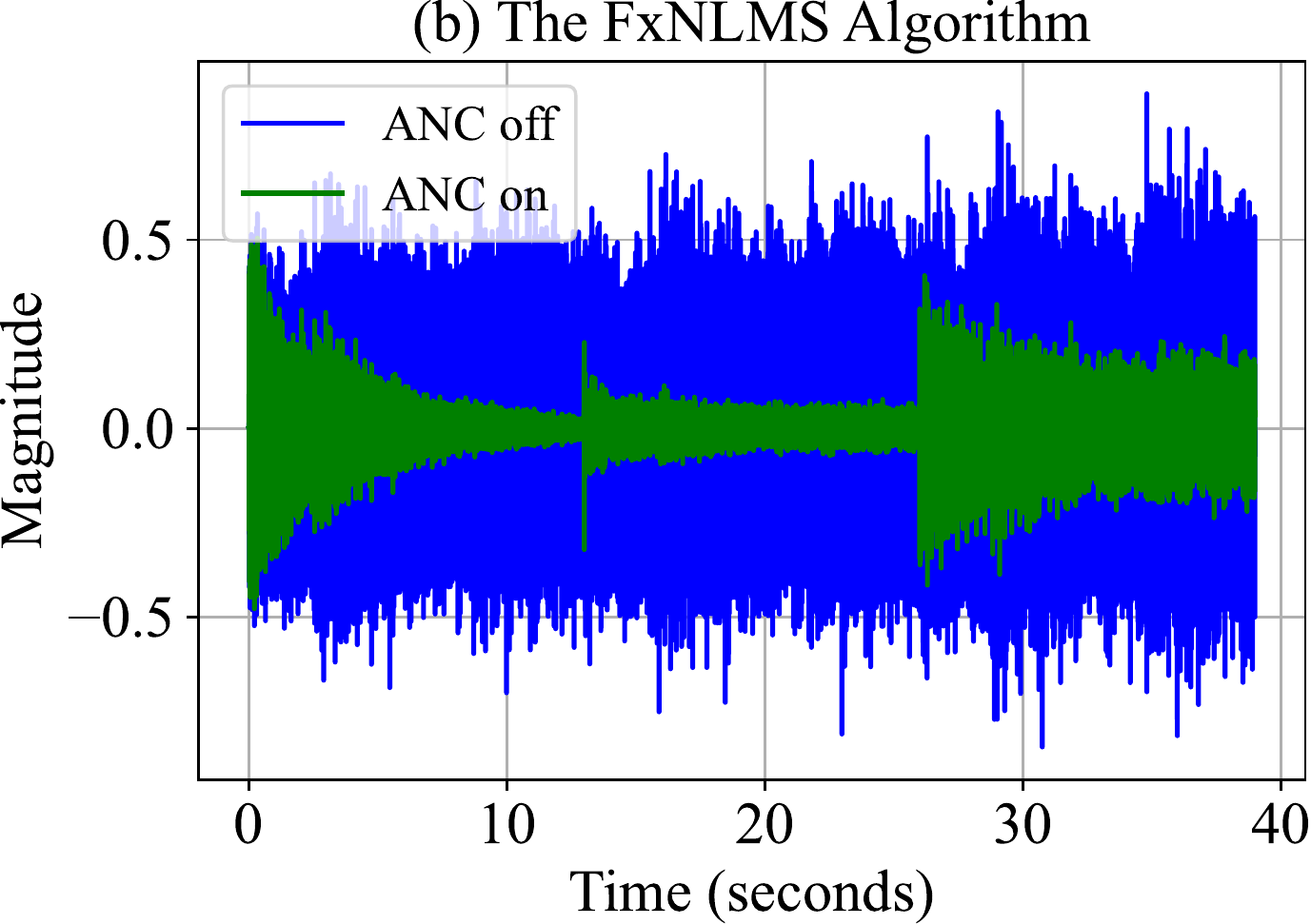}
}
\subfigure{
\includegraphics[width=0.44\linewidth]{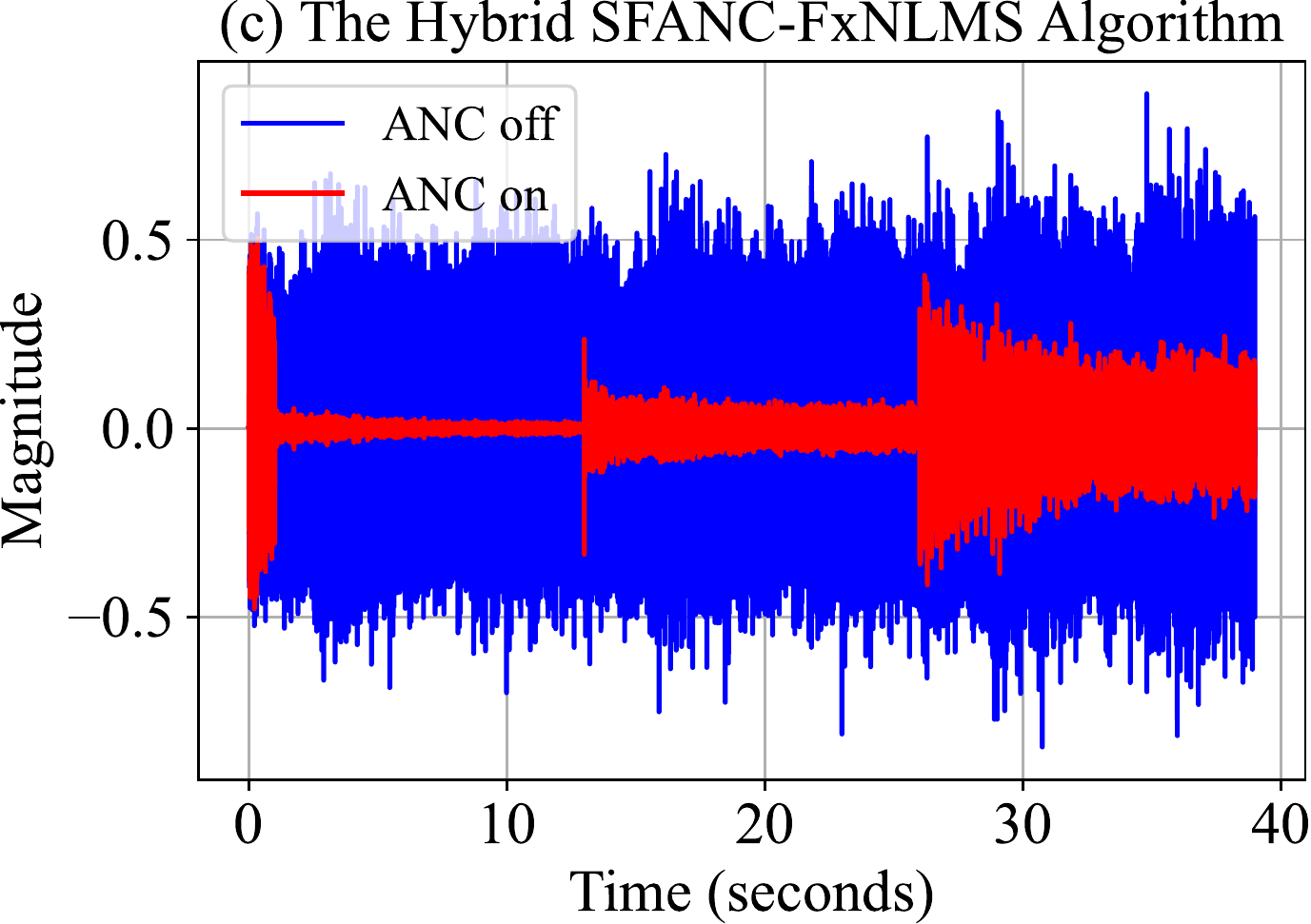}
}
\subfigure{
\includegraphics[width=0.44\linewidth]{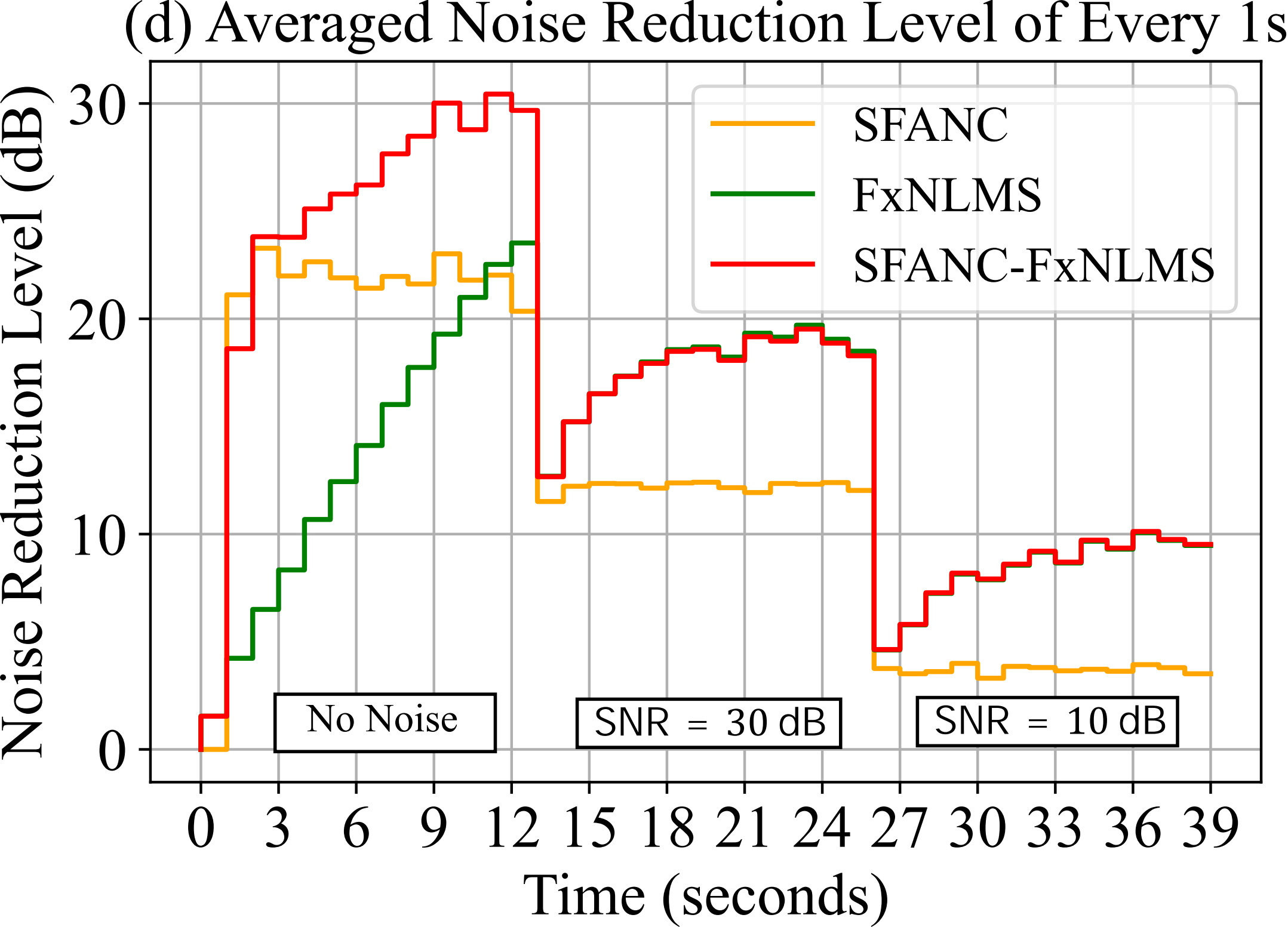}
}
\caption{(a)-(c): Error signals of different ANC algorithms, (d): Averaged noise reduction level of every 1 second, when dealing with a varying primary path.}
\label{Fig 7}\vspace*{-0.4cm}
\end{figure}

\subsection{Noise reduction performance on the varying primary path}
In this section, the robustness of different ANC algorithms are examined when dealing with a primary path that changes every $13$ seconds. The impulse response of the primary path is mixed with different levels of white noise to obtain additional signal-to-noise ratios (SNR) of $30$dB and $10$dB respectively as shown in Fig.~\ref{Fig 7}. The primary noise is the traffic noise.

According to Fig.~\ref{Fig 7}, in the initial period, the SFANC and hybrid SFANC-FxNLMS algorithm have better noise reduction performances over the FxNLMS algorithm. However, when the SNR of the primary path varies, SFANC can't perform as well as FxNLMS and hybrid SFANC-FxNLMS algorithm. Since the primary noise type remains the same, the selected pre-trained control filter is fixed in the SFANC algorithm, which may not be suitable on the varying primary path. The results indicate that the hybrid SFANC-FxNLMS algorithm is sufficiently robust to deal with the varying primary path.\vspace*{-0.2cm}

\section{Conclusion}
The proposed hybrid SFANC-FxNLMS algorithm effectively combines SFANC and FxNLMS to achieve faster noise reduction responses and lower steady-state errors. A lightweight 1D CNN operating at frame rate is used to automatically determine control filter coefficients. Also, the FxNLMS algorithm continues to update the coefficients adaptively at the sampling rate. Simulation results validate the good tracking capability and robustness of the hybrid SFANC-FxNLMS algorithm when dealing with varying noises and primary paths in dynamic noise control environments. The proposed hybrid algorithm is promising to be further explored for combining other fixed-filter and adaptive ANC algorithms.

\bibliographystyle{IEEEtran}
\bibliography{A}

\begin{thebibliography}{10}
\providecommand{\url}[1]{#1}
\csname url@samestyle\endcsname
\providecommand{\newblock}{\relax}
\providecommand{\bibinfo}[2]{#2}
\providecommand{\BIBentrySTDinterwordspacing}{\spaceskip=0pt\relax}
\providecommand{\BIBentryALTinterwordstretchfactor}{4}
\providecommand{\BIBentryALTinterwordspacing}{\spaceskip=\fontdimen2\font plus
\BIBentryALTinterwordstretchfactor\fontdimen3\font minus
  \fontdimen4\font\relax}
\providecommand{\BIBforeignlanguage}[2]{{%
\expandafter\ifx\csname l@#1\endcsname\relax
\typeout{** WARNING: IEEEtran.bst: No hyphenation pattern has been}%
\typeout{** loaded for the language `#1'. Using the pattern for}%
\typeout{** the default language instead.}%
\else
\language=\csname l@#1\endcsname
\fi
#2}}
\providecommand{\BIBdecl}{\relax}
\BIBdecl

\bibitem{1}
C.~N. Hansen, \emph{Understanding active noise cancellation}.\hskip 1em plus
  0.5em minus 0.4em\relax CRC Press, 2002.

\bibitem{elliot1994active}
S.~Elliot and P.~Nelson, ``Active noise control,'' \emph{Noise News
  International}, vol.~2, no.~2, pp. 75--98, 1994.

\bibitem{kuo1999active}
S.~M. Kuo and D.~R. Morgan, ``Active noise control: a tutorial review,''
  \emph{Proceedings of the IEEE}, vol.~87, no.~6, pp. 943--973, 1999.

\bibitem{kajikawa2012recent}
Y.~Kajikawa, W.-S. Gan, and S.~M. Kuo, ``Recent advances on active noise
  control: open issues and innovative applications,'' \emph{APSIPA Transactions
  on Signal and Information Processing}, vol.~1, 2012.

\bibitem{qiu2019introduction}
X.~Qiu, \emph{An introduction to virtual sound barriers}.\hskip 1em plus 0.5em
  minus 0.4em\relax CRC Press, 2019.

\bibitem{zhang2019active}
J.~Zhang, ``Active noise control over spatial regions,'' Ph.D. dissertation,
  The Australian National University (Australia), 2019.

\bibitem{chen2021secondary}
D.~Chen, L.~Cheng, D.~Yao, J.~Li, and Y.~Yan, ``A secondary path-decoupled
  active noise control algorithm based on deep learning,'' \emph{IEEE Signal
  Processing Letters}, vol.~29, pp. 234--238, 2021.

\bibitem{liebich2018signal}
S.~Liebich, J.~Fabry, P.~Jax, and P.~Vary, ``Signal processing challenges for
  active noise cancellation headphones,'' in \emph{Speech Communication; 13th
  ITG-Symposium}.\hskip 1em plus 0.5em minus 0.4em\relax VDE, 2018, pp. 1--5.

\bibitem{rivera2017evaluation}
P.~Rivera~Benois, P.~Nowak, and U.~Z{\"o}lzer, ``Evaluation of a decoupled
  feedforward-feedback hybrid structure for active noise control headphones in
  a multi-source environment,'' in \emph{INTER-NOISE and NOISE-CON Congress and
  Conference Proceedings}, vol. 255, no.~4.\hskip 1em plus 0.5em minus
  0.4em\relax Institute of Noise Control Engineering, 2017, pp. 3691--3699.

\bibitem{2}
S.~M. Kuo and D.~R. Morgan, ``Active noise control: a tutorial review,''
  \emph{Proceedings of the IEEE}, vol.~87, no.~6, pp. 943--973, 1999.

\bibitem{chang2016listening}
C.-Y. Chang, A.~Siswanto, C.-Y. Ho, T.-K. Yeh, Y.-R. Chen, and S.~M. Kuo,
  ``Listening in a noisy environment: Integration of active noise control in
  audio products,'' \emph{IEEE Consumer Electronics Magazine}, vol.~5, no.~4,
  pp. 34--43, 2016.

\bibitem{liebich2016active}
S.~Liebich, C.~Anem{\"u}ller, P.~Vary, P.~Jax, D.~R{\"u}schen, and
  S.~Leonhardt, ``Active noise cancellation in headphones by digital robust
  feedback control,'' in \emph{2016 24th European Signal Processing Conference
  (EUSIPCO)}.\hskip 1em plus 0.5em minus 0.4em\relax IEEE, 2016, pp.
  1843--1847.

\bibitem{shen2021wireless}
X.~Shen, D.~Shi, and W.-S. Gan, ``A wireless reference active noise control
  headphone using coherence based selection technique,'' in \emph{ICASSP
  2021-2021 IEEE International Conference on Acoustics, Speech and Signal
  Processing (ICASSP)}.\hskip 1em plus 0.5em minus 0.4em\relax IEEE, 2021, pp.
  7983--7987.

\bibitem{coker2019survey}
K.~Coker and C.~Shi, ``A survey on virtual bass enhancement for active noise
  cancelling headphones,'' in \emph{2019 International Conference on Control,
  Automation and Information Sciences (ICCAIS)}.\hskip 1em plus 0.5em minus
  0.4em\relax IEEE, 2019, pp. 1--5.

\bibitem{3}
R.~M. Reddy, I.~M. Panahi, and R.~Briggs, ``Hybrid fxrls-fxnlms adaptive
  algorithm for active noise control in fmri application,'' \emph{IEEE
  Transactions on Control Systems Technology}, vol.~19, no.~2, pp. 474--480,
  2010.

\bibitem{guo2020convergence}
J.~Guo, F.~Yang, and J.~Yang, ``Convergence analysis of the conventional
  filtered-x affine projection algorithm for active noise control,''
  \emph{Signal Processing}, vol. 170, p. 107437, 2020.

\bibitem{yang2018frequency}
F.~Yang, Y.~Cao, M.~Wu, F.~Albu, and J.~Yang, ``Frequency-domain filtered-x lms
  algorithms for active noise control: A review and new insights,''
  \emph{Applied Sciences}, vol.~8, no.~11, p. 2313, 2018.

\bibitem{4}
J.~C. Burgess, ``Active adaptive sound control in a duct: A computer
  simulation,'' \emph{The Journal of the Acoustical Society of America},
  vol.~70, no.~3, pp. 715--726, 1981.

\bibitem{shi2020activeNormalized}
D.~Shi, W.-S. Gan, B.~Lam, S.~Wen, and X.~Shen, ``Active noise control based on
  the momentum multichannel normalized filtered-x least mean square
  algorithm,'' in \emph{INTER-NOISE and NOISE-CON Congress and Conference
  Proceedings}, vol. 261, no.~6.\hskip 1em plus 0.5em minus 0.4em\relax
  Institute of Noise Control Engineering, 2020, pp. 709--719.

\bibitem{lu2021survey}
L.~Lu, K.-L. Yin, R.~C. de~Lamare, Z.~Zheng, Y.~Yu, X.~Yang, and B.~Chen, ``A
  survey on active noise control techniques--part i: Linear systems,''
  \emph{arXiv preprint arXiv:2110.00531}, 2021.

\bibitem{26}
Y.~Tsao, S.-H. Fang, and Y.~Shiao, ``Acoustic echo cancellation using a
  vector-space-based adaptive filtering algorithm,'' \emph{IEEE Signal
  Processing Letters}, vol.~22, no.~3, pp. 351--355, 2014.

\bibitem{22}
R.~Ranjan and W.-S. Gan, ``Natural listening over headphones in augmented
  reality using adaptive filtering techniques,'' \emph{IEEE/ACM Transactions on
  Audio, Speech, and Language Processing}, vol.~23, no.~11, pp. 1988--2002,
  2015.

\bibitem{5}
R.~Ranjan, T.~Murao, B.~Lam, and W.-S. Gan, ``Selective active noise control
  system for open windows using sound classification,'' in \emph{INTER-NOISE
  and NOISE-CON Congress and Conference Proceedings}, vol. 253, no.~6.\hskip
  1em plus 0.5em minus 0.4em\relax Institute of Noise Control Engineering,
  2016, pp. 1921--1931.

\bibitem{6}
C.~Shi, R.~Xie, N.~Jiang, H.~Li, and Y.~Kajikawa, ``Selective virtual sensing
  technique for multi-channel feedforward active noise control systems,'' in
  \emph{ICASSP 2019-2019 IEEE International Conference on Acoustics, Speech and
  Signal Processing (ICASSP)}.\hskip 1em plus 0.5em minus 0.4em\relax IEEE,
  2019, pp. 8489--8493.

\bibitem{11}
S.~Wen, W.-S. Gan, and D.~Shi, ``Using empirical wavelet transform to speed up
  selective filtered active noise control system,'' \emph{The Journal of the
  Acoustical Society of America}, vol. 147, no.~5, pp. 3490--3501, 2020.

\bibitem{7}
D.~Shi, W.-S. Gan, B.~Lam, and S.~Wen, ``Feedforward selective fixed-filter
  active noise control: Algorithm and implementation,'' \emph{IEEE/ACM
  Transactions on Audio, Speech, and Language Processing}, vol.~28, pp.
  1479--1492, 2020.

\bibitem{12}
D.~Shi, B.~Lam, and W.-S. Gan, ``A novel selective active noise control
  algorithm to overcome practical implementation issue,'' in \emph{2018 IEEE
  International Conference on Acoustics, Speech and Signal Processing
  (ICASSP)}.\hskip 1em plus 0.5em minus 0.4em\relax IEEE, 2018, pp. 1130--1134.

\bibitem{8}
Y.~LeCun, Y.~Bengio, and G.~Hinton, ``Deep learning,'' \emph{nature}, vol. 521,
  no. 7553, pp. 436--444, 2015.

\bibitem{21}
C.~Szegedy, W.~Liu, Y.~Jia, P.~Sermanet, S.~Reed, D.~Anguelov, D.~Erhan,
  V.~Vanhoucke, and A.~Rabinovich, ``Going deeper with convolutions,'' in
  \emph{Proceedings of the IEEE conference on computer vision and pattern
  recognition}, 2015, pp. 1--9.

\bibitem{sikora2021influence}
J.~Sikora, R.~Wagnerov{\'a}, L.~Landryov{\'a}, J.~{\v{S}}{\'\i}ma, and
  S.~Wrona, ``Influence of environmental noise on quality control of hvac
  devices based on convolutional neural network,'' \emph{Applied Sciences},
  vol.~11, no.~16, p. 7484, 2021.

\bibitem{zhang2021deep}
H.~Zhang and D.~Wang, ``Deep anc: A deep learning approach to active noise
  control,'' \emph{Neural Networks}, vol. 141, pp. 1--10, 2021.

\bibitem{9}
D.~Shi, B.~Lam, K.~Ooi, X.~Shen, and W.-S. Gan, ``Selective fixed-filter active
  noise control based on convolutional neural network,'' \emph{Signal
  Processing}, vol. 190, p. 108317, 2022.

\bibitem{10}
W.~Dai, C.~Dai, S.~Qu, J.~Li, and S.~Das, ``Very deep convolutional neural
  networks for raw waveforms,'' in \emph{2017 IEEE International Conference on
  Acoustics, Speech and Signal Processing (ICASSP)}.\hskip 1em plus 0.5em minus
  0.4em\relax IEEE, 2017, pp. 421--425.

\bibitem{23}
P.~A. Lopes and M.~Piedade, ``The kalman filter in active noise control,'' in
  \emph{INTER-NOISE and NOISE-CON Congress and Conference Proceedings}, vol.
  1999, no.~5.\hskip 1em plus 0.5em minus 0.4em\relax Institute of Noise
  Control Engineering, 1999, pp. 1111--1124.

\bibitem{24}
P.~A. Lopes and M.~S. Piedade, ``A kalman filter approach to active noise
  control,'' in \emph{2000 10th European Signal Processing Conference}.\hskip
  1em plus 0.5em minus 0.4em\relax IEEE, 2000, pp. 1--4.

\bibitem{kay1993fundamentals}
S.~M. Kay, \emph{Fundamentals of statistical signal processing: estimation
  theory}.\hskip 1em plus 0.5em minus 0.4em\relax Prentice-Hall, Inc., 1993.

\bibitem{25}
S.~van Ophem and A.~P. Berkhoff, ``Multi-channel kalman filters for active
  noise control,'' \emph{The Journal of the Acoustical Society of America},
  vol. 133, no.~4, pp. 2105--2115, 2013.

\bibitem{yang2020stochastic}
F.~Yang, J.~Guo, and J.~Yang, ``Stochastic analysis of the filtered-x lms
  algorithm for active noise control,'' \emph{IEEE/ACM Transactions on Audio,
  Speech, and Language Processing}, vol.~28, pp. 2252--2266, 2020.

\bibitem{19}
S.~Ioffe and C.~Szegedy, ``Batch normalization: Accelerating deep network
  training by reducing internal covariate shift,'' in \emph{International
  conference on machine learning}.\hskip 1em plus 0.5em minus 0.4em\relax PMLR,
  2015, pp. 448--456.

\bibitem{20}
M.~D. Zeiler, M.~Ranzato, R.~Monga, M.~Mao, K.~Yang, Q.~V. Le, P.~Nguyen,
  A.~Senior, V.~Vanhoucke, J.~Dean \emph{et~al.}, ``On rectified linear units
  for speech processing,'' in \emph{2013 IEEE International Conference on
  Acoustics, Speech and Signal Processing}.\hskip 1em plus 0.5em minus
  0.4em\relax IEEE, 2013, pp. 3517--3521.

\bibitem{14}
K.~He, X.~Zhang, S.~Ren, and J.~Sun, ``Deep residual learning for image
  recognition,'' in \emph{Proceedings of the IEEE conference on computer vision
  and pattern recognition}, 2016, pp. 770--778.

\bibitem{15}
S.~C. Douglas and T.-Y. Meng, ``An optimum nlms algorithm: Performance
  improvement over lms,'' in \emph{Acoustics, Speech, and Signal Processing,
  IEEE International Conference on}.\hskip 1em plus 0.5em minus 0.4em\relax
  IEEE Computer Society, 1991, pp. 2125--2126.

\bibitem{17}
D.~P. Kingma and J.~Ba, ``Adam: A method for stochastic optimization,''
  \emph{arXiv preprint arXiv:1412.6980}, 2014.

\bibitem{16}
X.~Glorot and Y.~Bengio, ``Understanding the difficulty of training deep
  feedforward neural networks,'' in \emph{Proceedings of the thirteenth
  international conference on artificial intelligence and statistics}.\hskip
  1em plus 0.5em minus 0.4em\relax JMLR Workshop and Conference Proceedings,
  2010, pp. 249--256.

\end{thebibliography}
\end{document}